\documentclass[twocolumn]{aastex631}
\usepackage{amsmath}
\usepackage[normalem]{ulem}

\renewcommand\altaffilmark{\textsuperscript{\dagger}}

\shorttitle{Evolution of reconnection flux}
\shortauthors{Maity et al.}

\begin{document}

\title{Evolution of reconnection flux during eruption of magnetic flux ropes}

\author[0009-0005-4347-9044]{Samriddhi Sankar Maity \textsuperscript{$\dagger$}}
\affiliation{Indian Institute of Astrophysics, Bangalore, India}
\affiliation{Joint Astronomy Programme and Department of Physics, Indian Institute of Science, Bangalore, India}

\author[0000-0002-0181-2495]{Piyali Chatterjee}
\affiliation{Indian Institute of Astrophysics, Bangalore, India}

\author[0000-0001-6457-5207]{Ranadeep Sarkar}
\affiliation{University of Helsinki, Finland}

\author[0009-0001-4457-4669]{Ijas S. Mytheen \textsuperscript{$\ddagger$}}
\affiliation{Indian Institute of Astrophysics, Bangalore, India}
\affiliation{Amrita School of Physical Science, Amrita Vishwa Vidyapeetham, Kollam, India}

\begingroup
\renewcommand{\thefootnote}{}
\footnotetext{\raggedright \hspace*{-1.2 em} \textsuperscript{$\dagger$} Now at Department of Physics \& Astronomy, Georgia State University, Atlanta, GA, USA and NASA Goddard Space Flight Center, Greenbelt, MD , USA}
\endgroup

\begingroup
\renewcommand{\thefootnote}{}
\footnotetext{\hspace*{-1.2 em} \textsuperscript{$\ddagger$} Now at  Department of Astronomy, Institute of Physics, Eötvös Loránd University (ELTE), Budapest, Hungary}
\endgroup

\begin{abstract}

Coronal mass ejections (CMEs) are powerful drivers of space weather, with magnetic flux ropes (MFRs) widely regarded as their primary precursors. However, the variation in reconnection flux during the evolution of MFR during CME eruptions remains poorly understood. In this paper, we develop a 3D magneto-hydrodynamic model using which we explore the temporal evolution of reconnection flux during the MFR evolution using both numerical simulations and observational data.
Our initial coronal configuration features an isothermal atmosphere and a potential arcade magnetic field beneath which an MFR emerges at the lower boundary. As the MFR rises, we observe significant stretching and compression of the overlying magnetic field beneath it. Magnetic reconnection begins with the gradual formation of a current sheet, eventually culminating with the impulsive expulsion of the flux rope. We analyze the temporal evolution of reconnection fluxes during two successive MFR eruptions while continuously emerging the twisted flux rope through the lower boundary. We also conduct a similar analysis using observational data from the Helioseismic and Magnetic Imager (HMI) and the Atmospheric Imaging Assembly (AIA) for an eruptive event. Comparing our MHD simulation with observational data, we find that reconnection flux play a crucial role in determination of CME kinematics. From the onset to the eruption, the rate of reconnection shows a monotonic variation with the acceleration. This simulation of a solar eruption provides important insights into the complex dynamics of CME initiation and progression.

\end{abstract}

\section{Introduction}
Coronal Mass Ejections (CMEs) are composed of clouds of magnetized plasma that are expelled from the Sun into the heliosphere due to sudden release of free magnetic energy stored in the twisted coronal magnetic field \citep{chen2017physics}.
They are of interest due to scientific and technological reasons since CMEs can drive interplanetary shocks that energize solar particles and cause significant space weather effects in the geospace. In-situ data obtained by satellites passing through the interplanetary CMEs (ICMEs) have established that typical ICMEs have the structure of highly twisted MFR \citep{wang2016twistimfr,hu2017grad}. Many CME models therefore incorporate a magnetic flux rope-- consisting of helical field lines twisting about a central axis in the corona as the basic underlying magnetic field structure for CME precursors \citep{titov1999topology, gibson2006prominences,duan2019pre-flare,liu2020magnetic, chen2017fluxropes}. Therefore, understanding their evolution in early phase is crucial for CME studies. 

Although is it widely accepted that MFRs constitute the core structure of the CMEs, it remains unknown wheather MFRs exist in the solar corona before CME initiation or forms during the eruption \citep{chen2011cmelivrev,patsourakos2020preeruptive}. Some opine that the MFR could exist prior to eruptions \citep{cheng2011magnetic}, although there is no consensus on how and where an MFR might form. An alternate  hypothesis is that the MFRs can bodily emerge from the below the photosphere \citep{fan2001twistedomega,martinez2008twisted,magara2004,archontis2009}. Yet another proposition is that the MFRs can be built directly in the corona via shearing of magnetic footpoints and reconnection prior to the eruption \citep{vanballegooijen1989,amari2003, aulanier2010, chatterjee2016repeatedflare}. MFRs can also form during the eruption \citep{Lynch2004, Wang2017}. Most eruptions, particularly those originating from solar active regions (ARs), occur along magnetic polarity inversion lines (PILs) within strong field regions. Furthermore, some flare-productive ARs exhibit relatively short time interval between successive eruptions, while displaying  very similar structure in the flare emissions and CME morphology. Such kind of event has been known as homologous eruptions \citep{Zhang2002}. Observations indicate that the evolution of the source regions of homologous events is often characterized by continuous shearing motion \citep{li2010L,romano2015,romano2018,Sarkar_2019}, sunspot rotation \citep{Regnier2006,zhang2008}, and flux emergence \citep{nitta2001,sterling2001,ranns2000,dun2007,xu2017}.

There exists several models to explain the slow build up and abrupt release of energy in solar eruptions \citep{Forbes2006,shibata2011,chen2011cmelivrev,schmieder2013,aulanier2014,janvier2015}. The two models widely used for homologous eruptions using magnetohydrodynamics (MHD) simulations are - the breakout reconnection models \citep{devore2008} and the slow tether cutting \citep{fan2010}. The breakout model invokes a multipolar magnetic configuration with a null point located above a central flux system which is sheared by photospheric motion. As the shearing increases, the magnetic reconnection begins at the null point above the newly forming flux rope. This reconnection process acts as a trigger for an eruption, When the eruption occurs, it removes the overlying magnetic field above the core that was previously restraining the core. After the eruption, the original magnetic structure is restored \citep{Antiochos1999,wyper2017,lynch2008}. Continuous shearing motions can result in the repetition of such mechanisms and give rise to multiple eruptions, which otherwise tend to be confined \cite{devore2008}. On the other hand, the slow tether-cutting model suggests that once a coronal flux rope is formed in the corona, its slow rise is governed by reconnection between the field lines of the twisted flux rope and the ambient field lines of the corona \citep{chen2000,archontis2008,fan2010}. The reconnection site usually lies below the rising flux rope. \cite{chatterjee2013} demonstrated repeated CMEs caused by a highly twisted flux tube emerging into the solar corona. In their simulation, a coronal flux rope partially erupted and reformed multiple times aided by partial internal reconnection between the legs of the flux rope. However, the time between eruptions was too short for the magnetic field to fully stabilize before the next eruption occurred.

Although magnetic reconnection can only be indirectly observed, it is a critical process in the solar corona that forms closed loops and energizes the plasma and particles leading to impulsively enhanced flare radiation. The observed correlation between the evolution of the CME kinematics and that of the flare X-ray fluxes suggests that the CME eruption is related to the reconnection \citep{zhang2001,temmer2008,bein2012,patsourakos2013}, consistent with CME initiation models \citep{chen2011cmelivrev}. Magnetic reconnection almost always occurs during a CME eruption yet it remains unclear whether reconnection initiates the eruption or a consequence. The morphological evolution of flare emission in the lower atmosphere has also been used to infer the reconnection process in the corona \citep{forbes1984,kopp1986}. Due to the difficulty in observing CME evolution and measuring CME acceleration in the low corona, many studies instead compare the CME velocity measured at a few solar radii after the peak acceleration phase and the total magnetic flux reconnected during the flare. Note that most of the studies have focused on an instant after the peak phase {\citep{Qiu2005,Pal2018,Gopalswamy2018}} and it is unclear whether the reconnection flux is related to the speed during the CME evolution. Numerical MHD simulations have proven to be powerful tools for reproducing the time-dependent, nonlinear evolution of 3D magnetic configurations and investigating the temporal changes in flux during the early phases of magnetic flux rope (MFR) eruptions. A key property characterizing solar flares is the amount of magnetic flux passing through the reconnection sheet beneath the MFR, commonly referred to as the reconnection flux \citep{Gopalswamy2017}. While the reconnected flux cannot be measured directly from observations of the corona, a quantitative relationship between the reconnection flux in the corona is the magnetic flux swept by the flare ribbon or the foot points of the magnetic flux rope \citep{kazachenko2017,Qiu2004}.

In this paper, we conduct a 3-dimensional compressible MHD simulation to study the time evolution of the reconnection flux and find the correlation with the speed of the magnetic flux rope during its early evolution. Next we aim at verifying the signatures of this correlation in observations. While, an ideal event required to estimate the reconnection flux correctly should be an on-disk event, the perfect event to calculate the ejection speed of the CME is instead a limb event. With the availability of multiple vantage points due to Stereo-A and SDO spacecraft provides an opportunity to check for particular events captured simultaneously by both spacecrafts. We analyse one of such rare events using coronagraph data from stero-A and EUV data from AIA instrument of SDO. The observational data and methods are described in \S~2, the numerical setup is presented in \S~3. The results are presented in \S~4 and finally we summarize our work.

\section{Numerical Model}
We numerically solve the complete magneto-hydrodynamic equations in three-dimensional spherical coordinates to investigate the early evolution of magnetic flux ropes in the solar corona. For this, we employ the Pencil Code \citep{pencilcode2021}, an open-source, highly modular, and MPI-parallelized code designed for compressible MHD flows \footnote{\url{http://pencil-code.nordita.org}}. Utilizing a sixth-order finite difference scheme and a third-order Runge-Kutta time-stepping method—among various other options provided by the Pencil Code—we solve the following set of compressible MHD equations.

\begin{equation}
    \frac{\mathrm{D} \ln \rho}{\mathrm{D} t} = - \mathbf{\nabla} \cdot \mathbf{U}
\end{equation}

\begin{equation}
    \frac{\mathrm{D} \mathbf{U}}{\mathrm{D} t} = -\frac{\mathbf{\nabla} p}{\rho} + \frac{GM_\odot}{r^2}\hat{\mathbf{r}} + \frac{\mathbf{J} \times \mathbf{B}}{\rho}+ \mathbf{F}_\mathrm{corr} + \rho^{-1} \mathbf{F}_\mathrm{visc}
\end{equation}

\begin{equation}
\frac{\partial \mathbf{A}}{\partial t} = \mathbf{U} \times \mathbf{B} - \eta \mu_0 \mathbf{J}
\end{equation}
\begin{equation}
\begin{split}
\rho c_v {T} \frac{\mathrm{D} \ln {T}}{\mathrm{D} t} =  -\rho c_v {T}(\gamma -1 ) \nabla \cdot \mathbf{U} + \eta\mu_0 \mathbf{J}^2 + 2 \rho \nu {\sf{S}}_{ij}^2 \\
+ \nabla \cdot \mathbf{q}_\mathrm{cond} - \rho^{2} \Lambda(T) + \mathcal{H} + {\mathbf \rho \zeta_\mathrm{shock} \left( \nabla \cdot \mathbf{U} \right)^2}
\end{split}
\label{eqn:temperature}
\end{equation}

In the above, $\mathbf{U}$ is the velocity field; $\mathbf{B}$ is the magnetic field; $\mathbf{A}$ is the magnetic potential; $\mathbf{J}$ is the current density; $\rho$, $p$ and $T$ are, respectively, the plasma density, pressure and temperature of the system; $c_v$ is the specific heat at constant volume. The viscous force is modelled as
\begin{equation}
\rho^{-1} \mathrm{F_{visc}} = \nabla \cdot (2 \rho \nu \mathbf{S}) + \nabla \left(\rho \zeta_{shock} \nabla \cdot \mathbf{U} \right)
\end{equation}
where $\nu$ is kinematic viscosity and $\mathrm{S}$ is the traceless rate-of-strain tensor. The coefficients $\zeta_{shock}$, represents shock viscosity which is given by 
\begin{equation*}
\zeta_{shock} = \nu_{shock} \left< max_{3} [(- \nabla \cdot \mathbf{U})_+] \right>
\end{equation*}
where $\left< max_{3} [(- \nabla \cdot \mathbf{U})_+] \right>$ means that at each grid point, a value is assigned corresponding to the maximum positive flow convergence ($-\nabla \cdot \mathbf{U} > 0$) within three neighboring grid points along each spatial dimension. The resulting values are then smoothed using a running mean over three neighboring grid points in each coordinate direction.

Additionally, we incorporate a semi-relativistic Boris correction $F_{corr}$ into the classical MHD momentum equation to reduce numerical diffusion and avoid the need for excessively small time steps in MHD simulations, achieved by using an artificially reduced speed of light \citep{chatterjee2020}.

\begin{figure}[!t]
\centering
  \includegraphics[width=\linewidth]{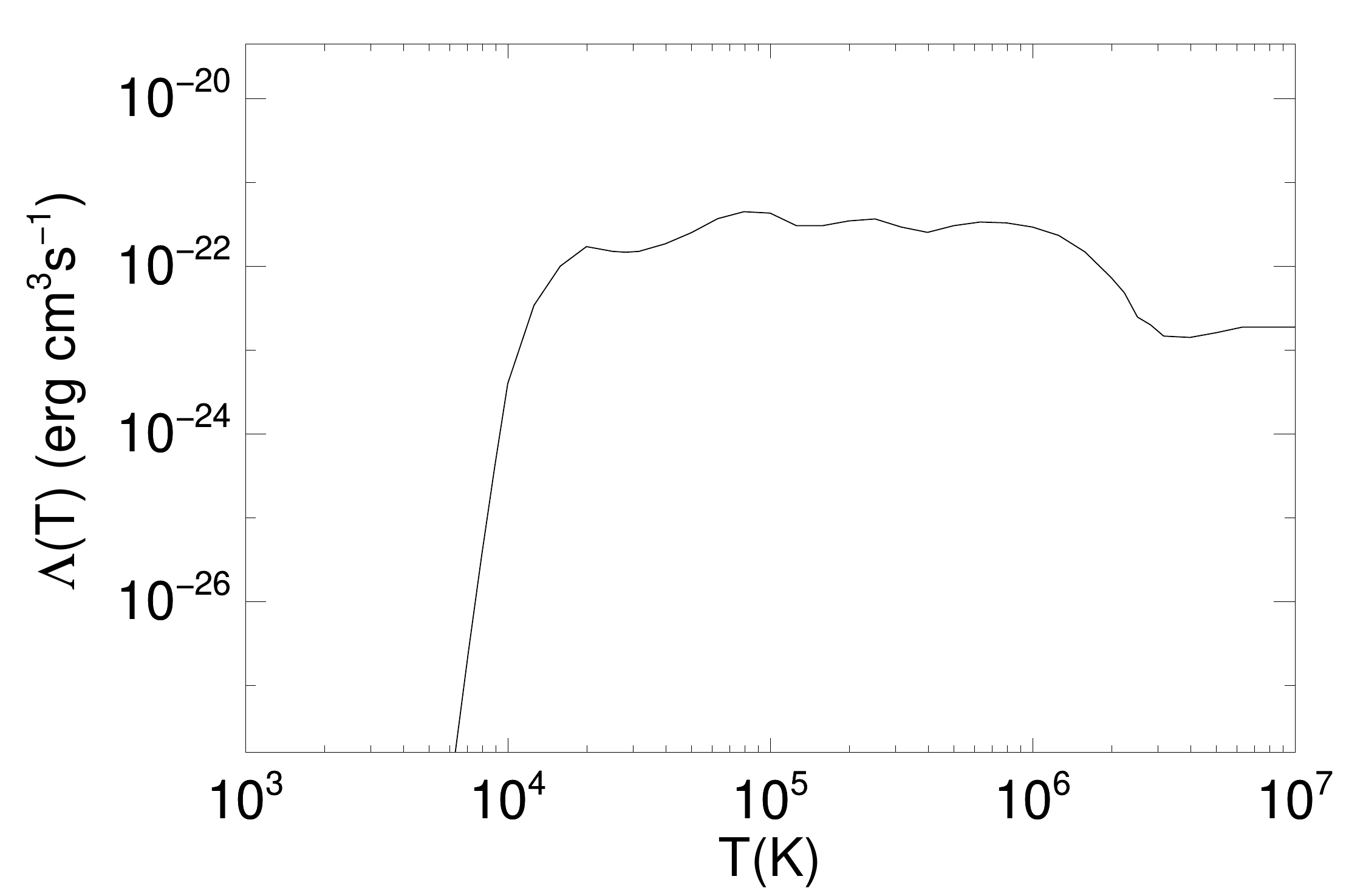}  
  \caption{The radiative cooling function in used in this work is a modified version of \cite{Cook1989}}
  \label{fig:radcool}
\end{figure}

We include radiative cooling in our model, represented in equation (\ref{eqn:temperature}), where $\Lambda(T)$ denotes the radiative loss function as described by \citep{Cook1989}. The cooling function is plotted in Figure (\ref{fig:radcool}). Furthermore, we use a simplified coronal heating function that varies only with height, following an exponential decay.
\begin{equation}
\mathcal{H} = \frac{F}{L_H} \frac{R_{\odot}^2}{r^2} \mathrm{exp}\left\{-(r-R_{\odot})/L_H\right\}
\end{equation}
where the input energy flux density is $F=9.74 \times 10^5 \, \mathrm{erg \, cm^{-2} \, s^{-1}}$ and the decay length is $L_H = 1.948 \times 10^{10} \, \mathrm{cm}$.

We have also included field-aligned Spitzer thermal conduction in our model.
Here we have used the hyperbolic diffusion equation approach instead of directly including Spitzer conduction in equation (\ref{eqn:temperature}), similar to \cite{chatterjee2020}.

Let $\mathbf{q}_{\mathrm{cond}}$ represents the solution of the non-Fickian transport equation and $\mathbf{q}_{\mathrm{sp}}$ denotes the conduction flux according to the Spitzer model. We solve the following equation for the heat flux $\mathbf{q}_{\mathrm{cond}}$.

\begin{equation}
\frac{\partial \mathbf{q}_\mathrm{cond}}{\partial t} = - \frac{\mathbf{q}_\mathrm{cond} - \mathbf{q}_{\mathrm{sp}}}{\tau_{sp}} + \beta (dr \cdot \nabla)^6 \mathbf{q}_\mathrm{cond}
\end{equation}

where $\mathbf{q}_{\mathrm{sp}} = K_{sp} T^{5/2} \hat{\mathbf{b}}(\hat{\mathbf{b}} \cdot \nabla T)$, $\hat{\mathbf{b}}$ denotes the unit vector along the field direction and $K_{sp} = 10^{-6}$ $\mathrm{erg \, K^{-7/2} \, cm^{-1} \, s^{-1}}$. $\tau_{sp}$ represents a finite timescale for $\mathbf{q}_{\mathrm{cond}}$ to evolve toward the Spitzer heat flux and is set to 0.1 s. Our time step varies between 0.1 - 0.3 ms. While our setup includes most of the relevant physics, solar wind is not included at the moment.

The computational domain consists of a spherical wedge with an extent $R_{\odot}< r < 6R_{\odot}$, $5\pi/12 < \theta < 7\pi/12$ and $-\pi/9.6 < \phi < \pi/9.6$. It is resolved by a grid of $512 \times 288 \times 160$, which is non uniform in $r$ but uniform in $\theta$ and $\phi$. The  grid spacing in $r$ is $dr = 0.002R_\odot$\, at the lower boundary which gradually increases in a logarithmic manner reaching $dr=0.003 R_\odot$\ at the upper boundary. Although the lower boundary in our setup is set at $R_{\odot}$, it does not represent the photosphere; instead, it serves as the lower coronal boundary.

\begin{figure}[!b]
\centering
  \includegraphics[width=\linewidth]{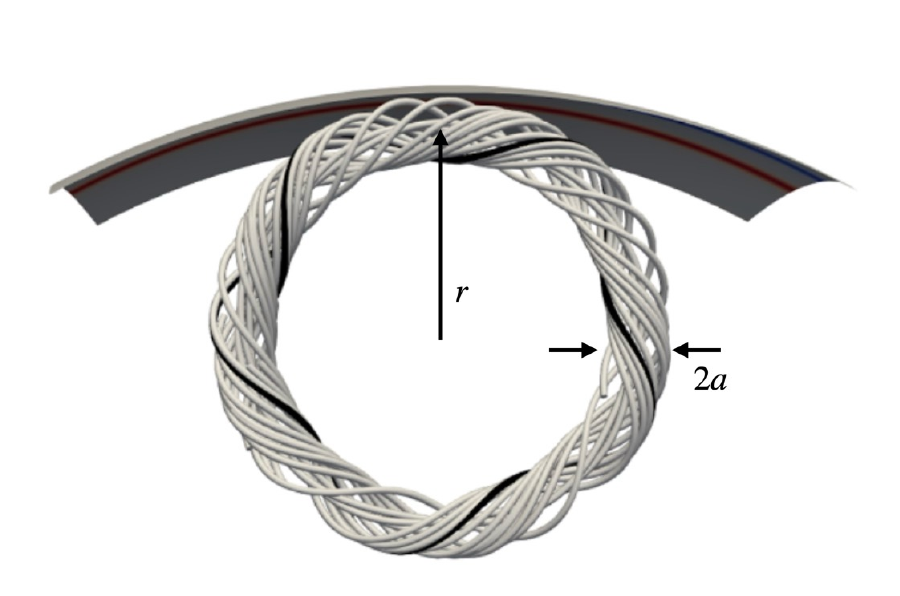}  
  \caption{The twisted torus before emerging through the bottom boundary. The major ($r$) and minor ($a$) radius of the torus is $0.25R_{\odot}$ and $0.042R_{\odot}$. The number of turns of one of the field lines is shown in black.}
  \label{fig:twisted_torus}
\end{figure}

\begin{figure*}
\centering
  \includegraphics[width=\linewidth]{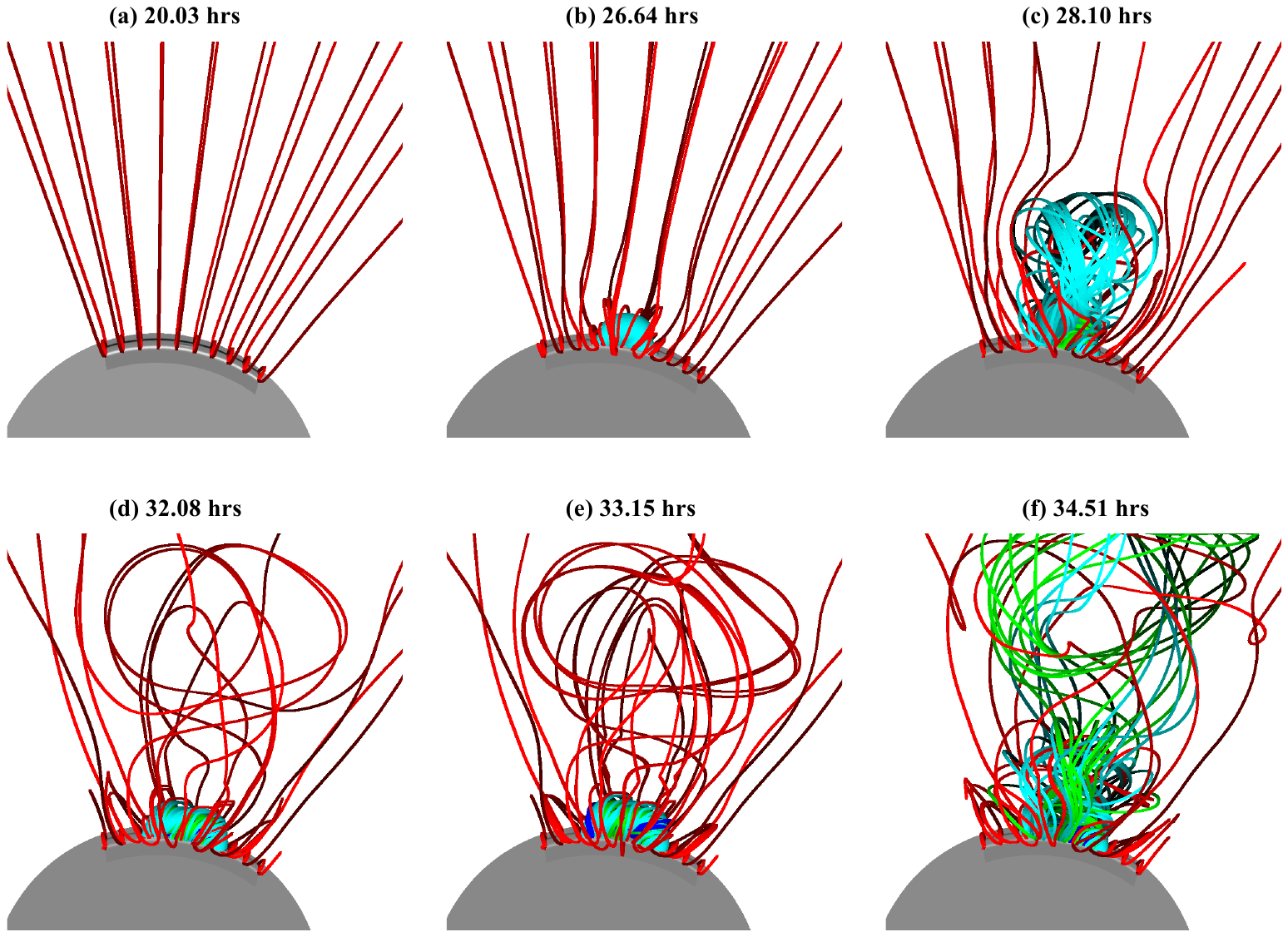}  
  \caption{The 3D evolution of the magnetic field of the twisted flux rope emerging into the corona at the specified times (in hours). Red field lines have footpoints in the ambient arcade, while blue (core), green (middle), and cyan (periphery) field lines originate from the emerging flux region at different distances from the flux rope axis. An animation of this evolution is available in the online article with a running time of 36\,s. The period covered by the animation (in solar hours) spans from $t=19.65$\,hrs to $t=40.25$\,hrs from the start of the simulation. The flux rope starts emerging into the corona at $t=20.3$\,hrs. }
  \label{fig:evolution}
\end{figure*}

We consider a high-temperature coronal plasma as an ideal gas with an adiabatic constant of $\gamma = 1.66$ . Initially, the domain is assumed to be in the state of a hydrostatic equilibrium at a uniform temperature of $T_0 = 10^6$\,K with a density stratification given by
\begin{equation}
\rho = \rho_0 \mathrm{exp}\left\{-\frac{R_{\odot}}{H_{p_0}}(1 - \frac{R_{\odot}}{r})\right\}
\end{equation}
Here, $\rho_0 = 1.0\times 10^{-15}$\,g cm$^{-3}$ is the initial density and the initial pressure scale height is given by $H_{p_0} = c_{s_0}^2/g_r$, where $c_{s_0}$ denotes the initial sound speed and $g_r$ represents the gravitational acceleration at the solar surface.
The initial atmosphere consists of a pre-existing potential field with arcade like geometry as given in \cite{fan2012} (see their Equation~12 and 13).

At the lower boundary, we impose an electromotive force given by 
\[
\mathbf{E}\mid_{r=R_{\odot}} = - \frac{1}{c} \mathrm{\mathbf{v}_0} \times \mathbf{B}_\mathrm{torus}(R_{\odot}, \theta, \phi, t)
\]
which bodily transports the twisted torus radially into the domain. The major and minor radius of the torus are $0.25R_{\odot}$ and $0.042R_{\odot}$ respectively (see Fig.~\ref{fig:twisted_torus}) and with a field line twist rate of 0.068\,rad\,Mm$^{-1}$. The twist rate indicates the field line winds around the tube axis per unit length along the axis. The flux tube is initially located at $r_0=0.705{R_{\odot}}$, thus the torus is initially entirely below the surface and moves bodily upward toward the lower boundary at constant speed, until it reaches a height $r_\mathrm{stop}$ when the emergence is stopped and a part of the torus is still inside of the computational domain. Here $r_0$ denotes the position of the center of the torus at the beginning of the injection whereas $r_\mathrm{stop}$ represents the position of the center of the torus when the emergence stopped. The velocity field at the lower boundary is specified to be uniform in the region where the emerging torus intersects the lower boundary and is zero everywhere else. The imposed emergence speed, $v_0$ = 2 $\mathrm{km\,s^{-1}}$, which is much smaller than the Alfven speed $v_A$ = 1.69 $\mathrm{Mm\,s^{-1}}$ to ensure that the emerging flux rope is allowed to evolve quasi-statically during the flux emergence phase at the lower boundary. The method for introducing a twisted torus structure into the domain quasi-statically from the lower boundary using an electromotive force is well-known and has been used by \cite{fan2009, fan2010, fan2012, fan2017} and \cite{chatterjee2013}. It is a technique that gradually introduces the flux rope into the corona. The corona is expected to be in a force-free equilibrium state at all times. We avoid adding the flux rope suddenly into the corona as that may lead to numerical perturbations that disturb the existing equilibrium and perhaps trigger an unphysical eruption. It is desirable to discourage this possibility and to be sure that the loss of equilibrium is indeed due to MHD instabilities like torus and helical kink rather than numerical. When the emergence is stopped, the velocity at the lower boundary is set to zero, with no inflows or  outflows and footpoints are rigidly anchored. For the $\theta$ boundaries, we assume a non-penetrating stress-free boundary for the velocity field and perfectly electrically conducting walls for the magnetic field. The $\phi$ boundaries are periodic. For the top boundary, we use a simple outward extrapolating boundary condition that allows plasma and magnetic field to flow through.

In this simulation, we drive the emergence of the torus until $r_\mathrm{stop}=0.85 R_{\odot}$ to study the rate of change of reconnection flux during the evolution of the twisted flux rope in the corona which produces homologous CMEs.

\section{Results}

Our simulation produces two homologous CME eruptions; each of the eruption characterized by an impulsive increase of the kinetic energy and corresponding release of the magnetic energy. We first briefly discussed the initiation and evolution of the eruptions and then analyze the velocity, acceleration and reconnection flux of different CME eruptions in the simulation. 
Note that the magnetic reconnection in our model using MHD approximation is of numerical origin, we cannot resolve length scales smaller than the Larmor radius of the proton or the mean free path. However, our goal here is to find if any relation exists between the rate of change of reconnection flux and the acceleration of the ejecta in our simulations.

\subsection{Synthetic eruptions}

\begin{figure}
\centering
  \includegraphics[width=\linewidth]{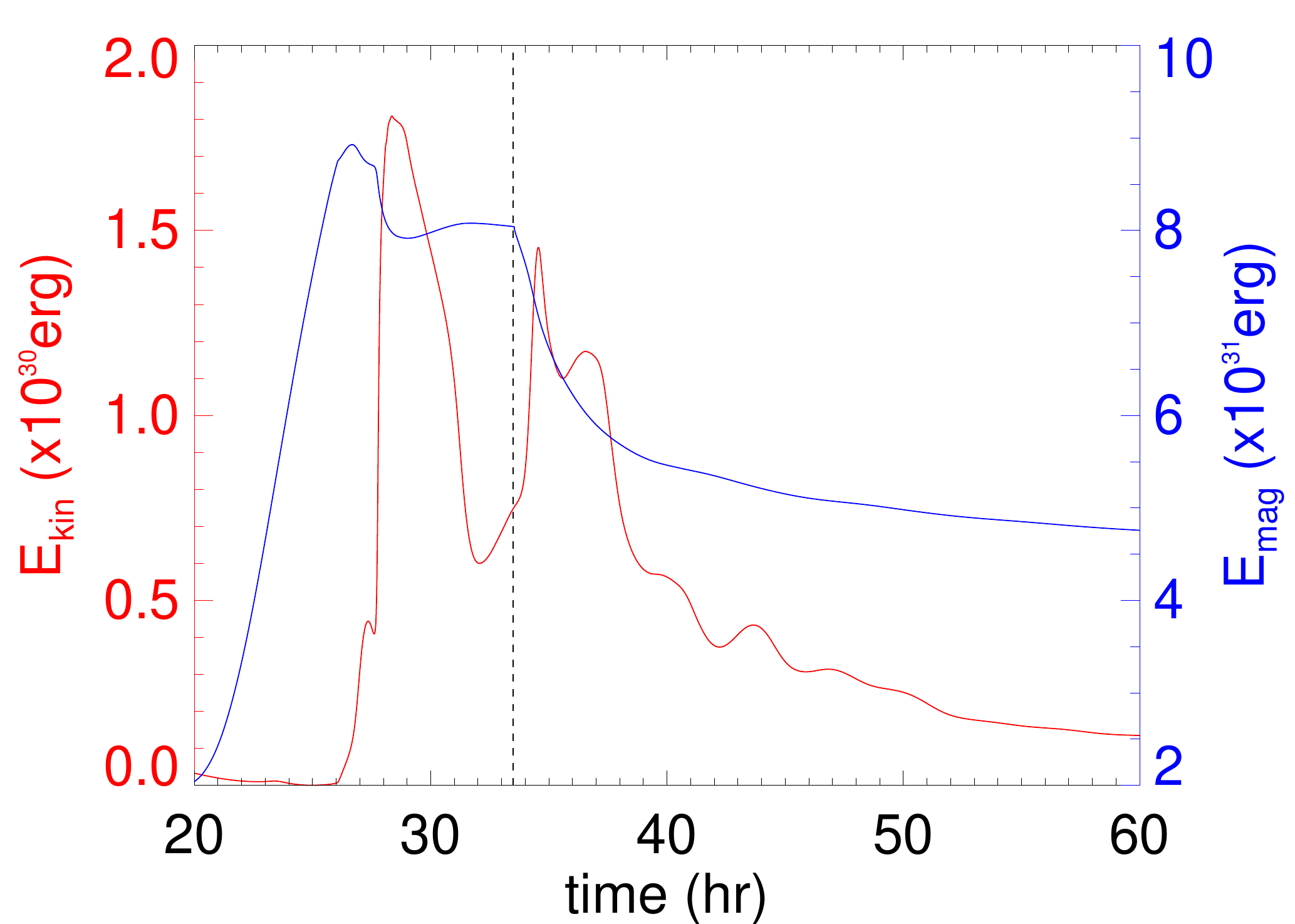}  
  \caption{Total kinetic energy $E_{kin}$ (red) and total magnetic energy $E_{mag}$ (blue) as a function of time. The dashed vertical line represents the time when the flux emergence stops.}
  \label{fig:energies}
\end{figure}

\begin{figure*}[!t]
\centering
  \includegraphics[width=\linewidth]{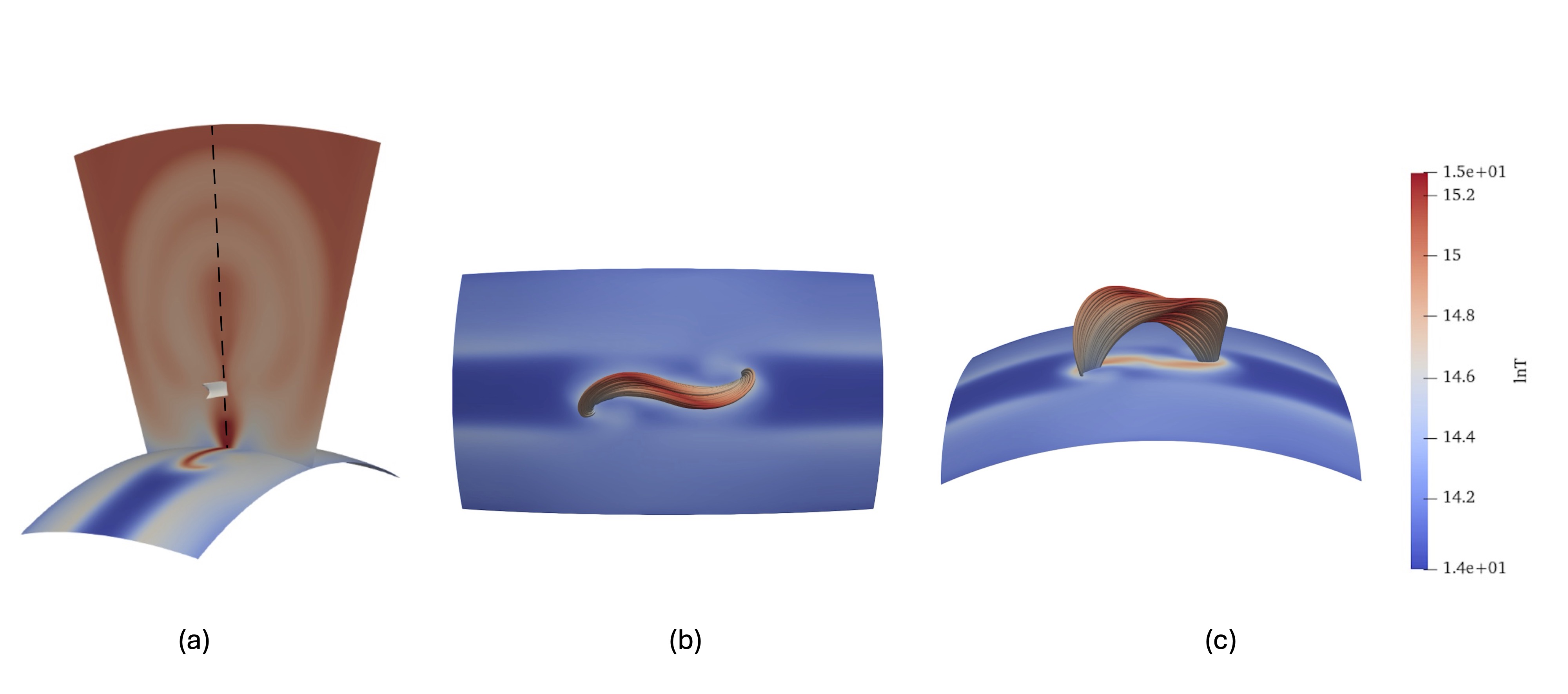}  
  \caption{Identification of current sheet and field lines of the flux rope from different viewing angles. The current sheet (in white) traced from the temperature 
  isosurface with a value $\log_{10} T = 6.6$ shown in panel (a). 
  The height of the current sheet iso-surface (white) from lower boundary is about $1.4R_\odot$. The field lines passing through such sheets gives the reconnection flux. The sigmoid fieldlines are shown in panels (b) and (c). The time in all these panels are taken at $t=27.32$\,hours.}
  \label{fig:CS}
\end{figure*}

The simulation begins with a magnetic flux rope (MFR) emerging from the lower boundary and pushing into a pre-existing coronal potential arcade field. The normal magnetic flux distributions at the lower boundary are represented by bipolar bands $B_r(R_{\odot}, \theta, \phi) = B_s(\theta)$ where the $B_s(\theta)$ denotes the potential arcade field at the lower boundary, as shown in Figure (\ref{fig:evolution}). The emergence process is halted once a specified amount of twisted flux is driven into the corona. Although the imposed emergence of a twisted flux rope through the lower boundary may not perfectly reflect real-sun conditions, it serves as a means to achieve a sequence of near-force-free coronal flux rope equilibria with increasing amounts of locally detached, twisted flux. In all cases, we observe the development of current layers with a sigmoid morphology beneath the flux rope. Magnetic reconnection within these current layers continues to add flux to the flux rope, even though the total magnetic energy gradually decreases due to reconnections. This continuous addition of flux facilitates successive eruptions.

\begin{figure}[!t]
    \centering
    \includegraphics[width=\linewidth]{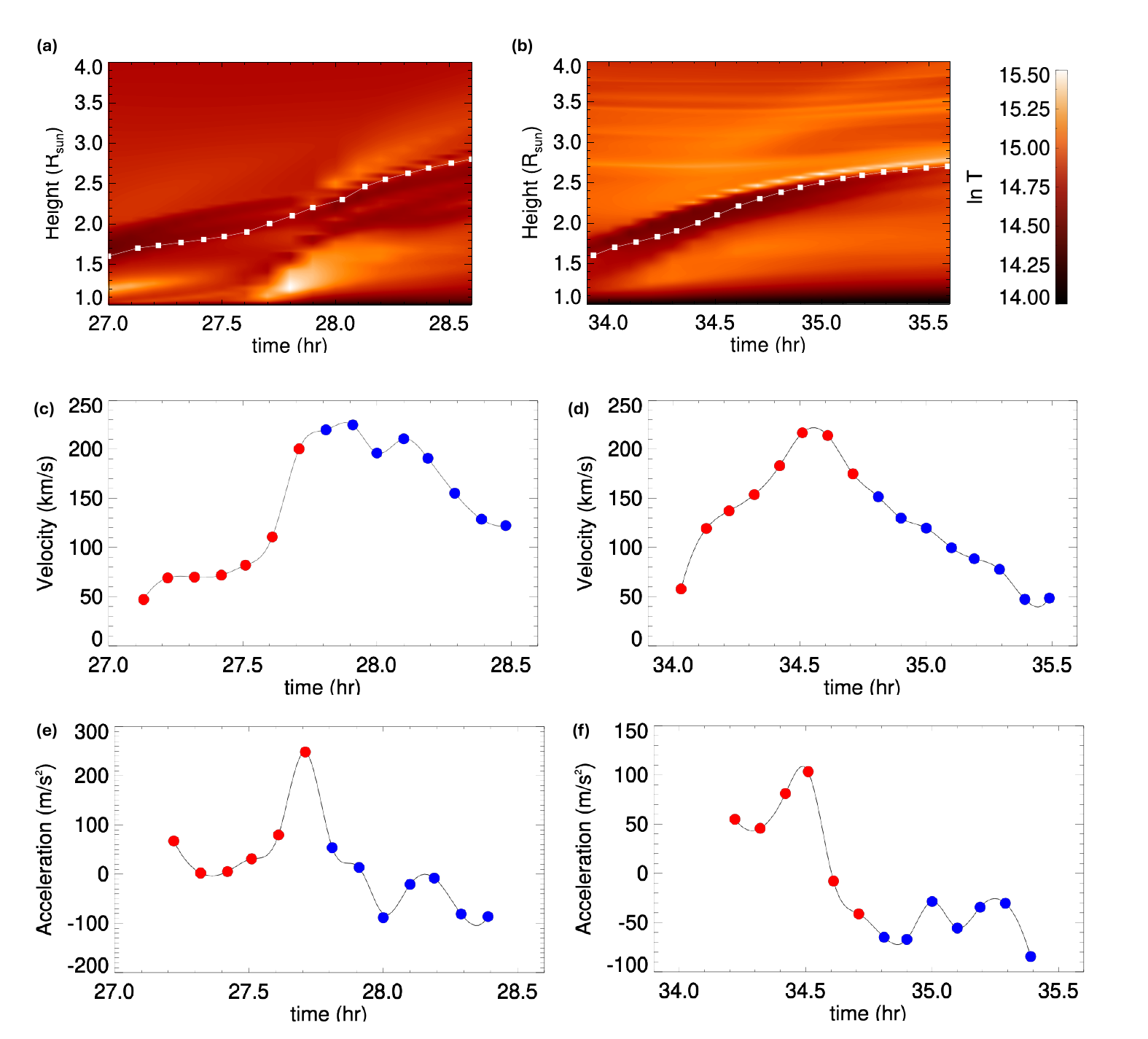}
    \caption{(a) Height time profile of the core of the flux rope before and after the first eruption at $t=27.9$\,hours. The height is calculated by tracking the dark region (flux rope) in the central meridional plane. (b) Same as (a), but for the second eruption at $t=34.7$\,hours. Time evolution of velocity, (c)-(d), and acceleration, (e)-(f), for both the eruptions in the simulation. The red (blue) circles are used to represent simulation data before (after) the eruption.}
    \label{fig:combined_sim}
\end{figure}

The first and second eruptions are triggered at $t= $ 27.7 hr and 34.8 hr. Figure~\ref{fig:energies} illustrates the evolution of kinetic and magnetic energies for both eruptions. The time at which the kinetic energy reaches its peak is designated as the eruption time.

The panels (c) and (d) of Figure \ref{fig:combined_sim} illustrate the radial velocity of the flux rope during both eruptions. These velocities were determined by tracking the flux rope axis throughout their evolution. The height-time plot is produced by tracking the dark region in the  $r-\theta$  plane of the temperature profile, which represents the MFR. By following this dark region consistently, we determine the MFR’s velocity during the flux rope’s evolution. The first ejection has a peak radial velocity of $224 \, \mathrm{km \, s^{-1}}$, which is slightly higher than the $213 \, \mathrm{km \, s^{-1}}$ velocity of the second CME. This observation is understood by the time versus kinetic energy plot, which shows the second CME registered lower kinetic energy than its precursor. The initial eruption occurs at a radial distance of $1.9 R_\odot$, while the second eruption takes place at $1.8 R_\odot$. Due to the lower velocity of the second eruption, there is no interaction between the two CMEs. An interesting observation is that the velocity starts to decrease after the eruption.

The reconnection flux (RC flux) passing through the current sheet is given by
\[
\phi_{RC} = \int B_n \, dS = \int B_r \, dA
\]
where \( B_n \) represents the magnetic flux normal to the elemental area \( dS \) of the current sheet, and \( B_r \) is the radial magnetic field in the ribbons, which are the footpoints of newly reconnected magnetic field lines near the lower boundary shown in Fig.~\ref{fig:CS}. 
The term \( dA \) denotes the elemental area of these footprints. 
Thus reconnection flux is determined by calculating the magnetic flux 
swept by the field lines passing near the lower boundary. Direct measurement of \( B_n \) and \( dS \) within the current sheet is not feasible, even in simulations, due to the 
intricate structure of the reconnection sheet. However, \( B_r \) and \( dA \) are relatively straightforward to calculate, even for a spherical wedge-shaped domain. Therefore, this approach provides an indirect but well-defined measure of the magnetic flux 
passing through the reconnection sheet. This method allows for an effective evaluation of the magnetic reconnection process by focusing on the 
measurable quantities at the footpoints, offering valuable insights into the dynamics of the reconnection events.

\begin{figure}
\centering
  \includegraphics[width=\linewidth]{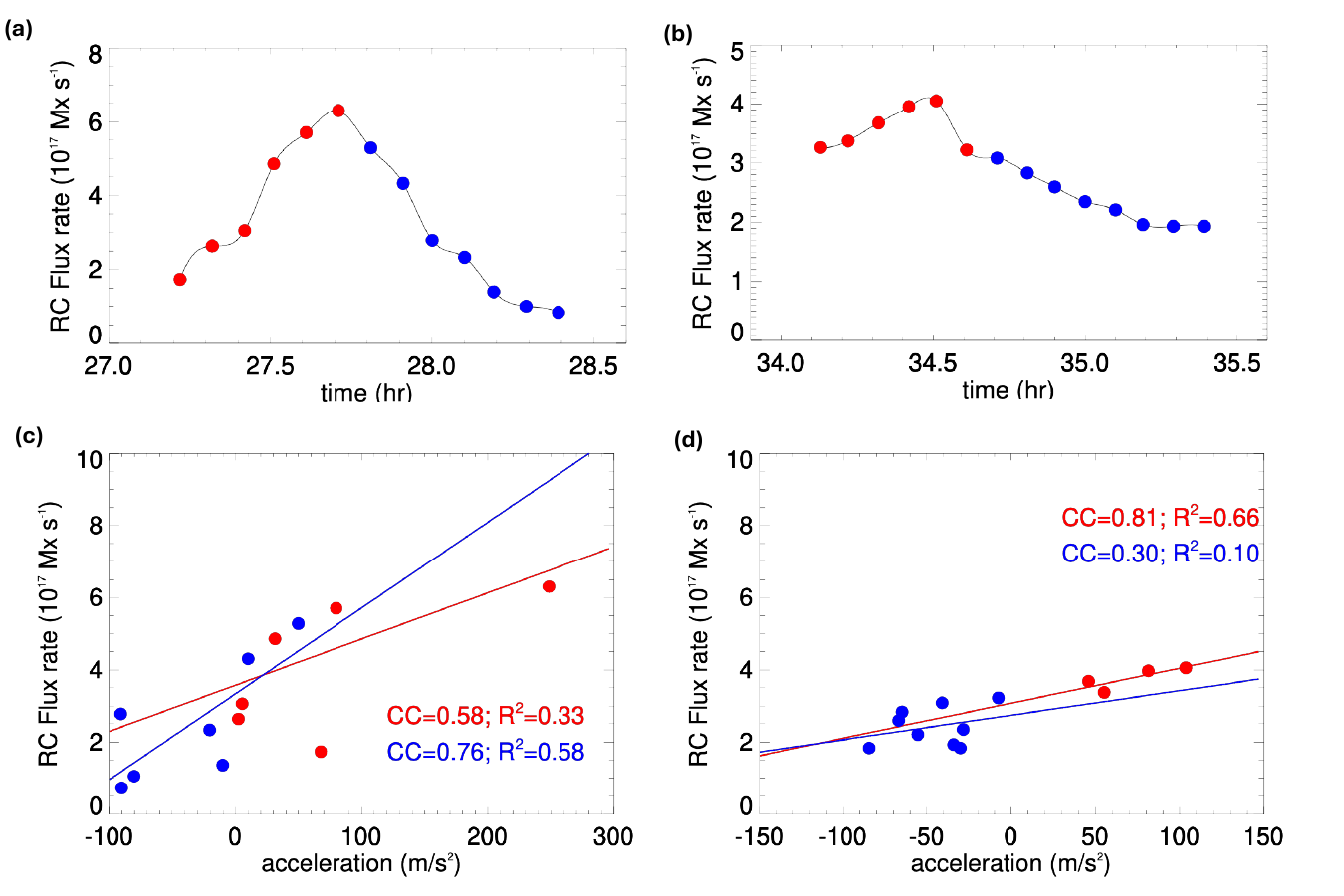}  
  \caption{Time evolution of reconnection rate (a)-(b). The reconnection flux is calculated by the magnetic flux passing through the footpoints of the fieldlines as well as through the current sheet shown in Fig.~\ref{fig:CS}a. The red (blue) circles are used to represent simulation data before (after) the eruption. The smoothing is done by cubic spline. (c) CME reconnection rate vs acceleration plot for the first eruption. The red (blue) line is the linear fit to the points before (after) the eruption, respectively. (d) Same as (c) but for the second eruption.}
  \label{fig:fluxrate}
\end{figure}

\begin{figure*}
\centering
\includegraphics[width=\linewidth]{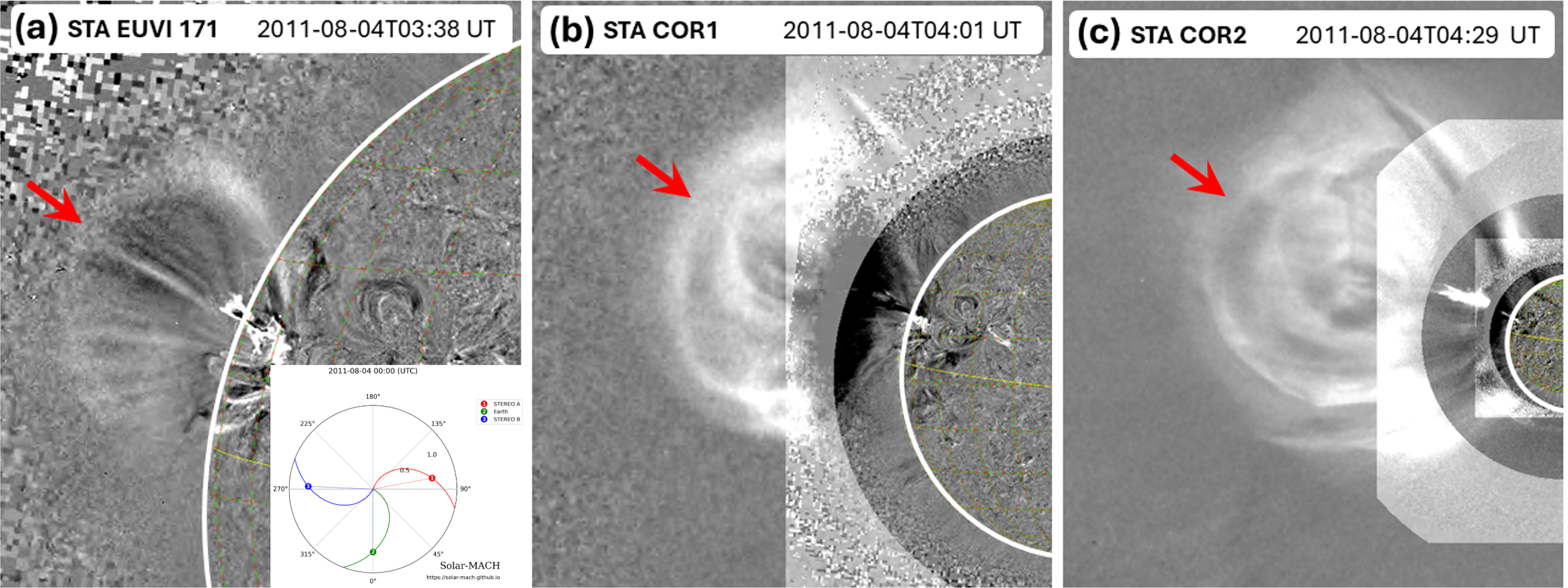}  
\caption{Different phases of the CME eruption on 2011 August 04 as observed by STEREO-A. (a) Base difference image of the Sun in STEREO-A EUVI 171 \AA\ . Location of STEREO spacecraft and Earth on 2011 August 4. shown in inset. (b) COR1 base-difference image superimposed with STEREO-A EUVI 171 \AA\ \. (c) Superimposed base-difference images of STEREO-A EUVI 171 \AA\ , COR1 and COR2. The red arrows in panels (a) and (b) indicate the CME leading edge. The images are plotted using JHelioviewer (\url{https://www.jhelioviewer.org/}).} 
\label{st_obs}
\end{figure*}

\begin{figure}[b]
\centering
\includegraphics[width=\linewidth]{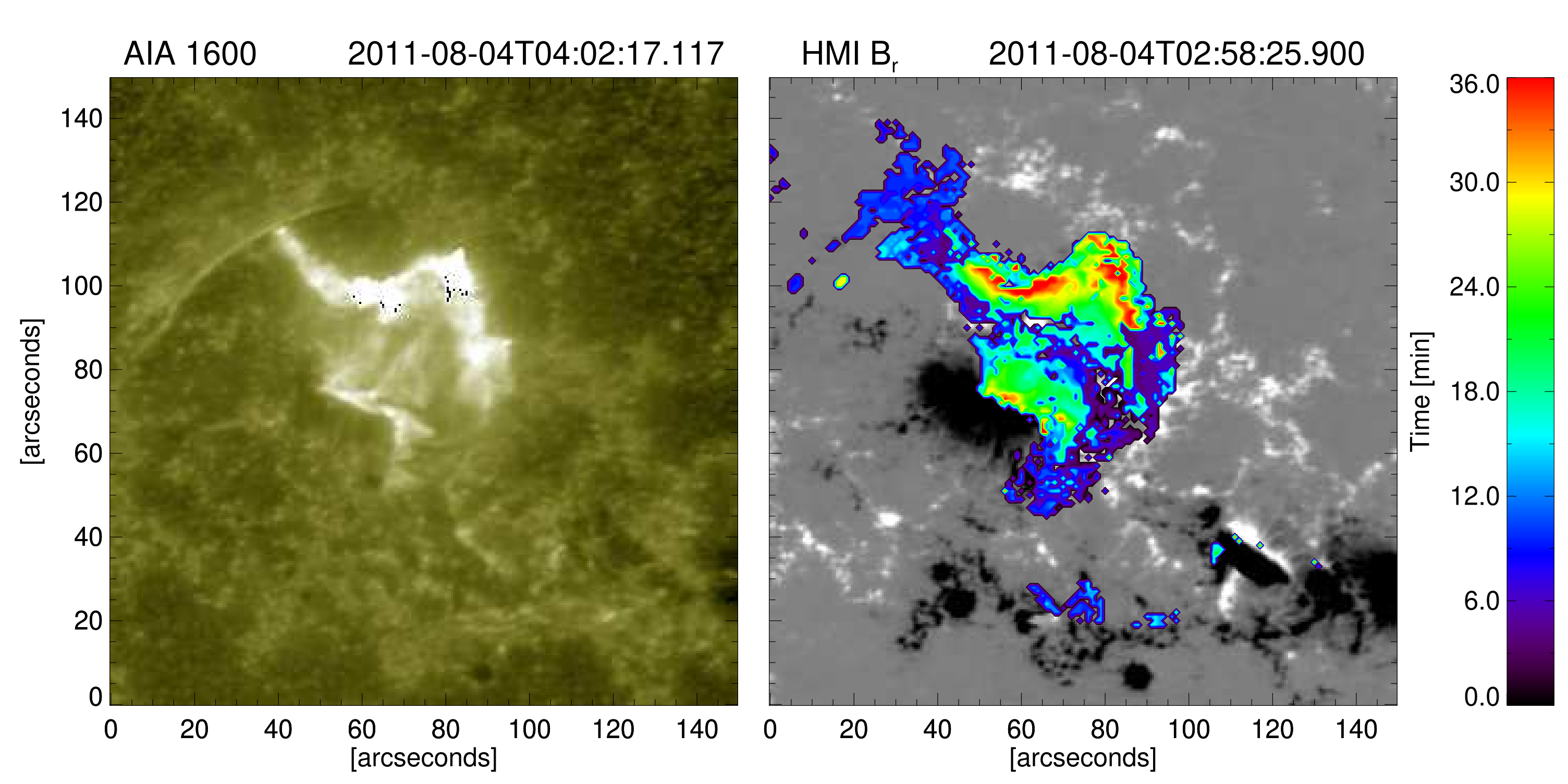}  
\caption{Left panel: Flare associated brightening observed in AIA 1600 \AA\ channel during the flare occurred in AR 11261. Right panel: The associated HMI $B_{r}$ magnetic field (gray color scale with saturation value $\pm$500 Gauss) and the temporal evolution of the flare ribbons from 03:41 UT to 04:17 UT on 2011 August 04 .}\label{rec_mag}
\end{figure}

The panels (a) and (b) of Figure \ref{fig:fluxrate} depicts the changes in reconnection rate (RC flux rate) over time for both eruptions. We monitored the hottest region beneath the flux rope in the $r \textrm{-} \theta$ plane both before and after the eruption as shown in Figure~\ref{fig:CS}). The peak reconnection rate (RC flux rate) for the first and second eruptions are $6.49 \times 10^{17}$\,$\mathrm{Mx \, s^{-1}}$ and $4.10 \times 10^{17}$\,$\mathrm{Mx \, s^{-1}}$,  respectively. The panels (a) and (b) of Figure~\ref{fig:fluxrate} indicate that the reconnection rate decreases after reaching its peak value, with the eruptions occurring near the maximum. This is similar to the observational data shown in Fig.~\ref{fig:combined_obs}b. After an eruption occurs, the reconnection sheet breaks down, making it challenging to track the field lines passing through the current sheet (CS).

\subsection{Observed eruption}
We now focus on finding signatures of such a correlation, if any, between reconnection flux and CME speed in solar observations. We study the temporal evolution of an M-class flare (SOL2011-SOL2011-08-04T03:41) which occurred in AR 11261 during 2011 August 4, and which had an associated CME. During this period, STEREO-A and STEREO-B were positioned at longitudinal separations of 101$^{\circ}$ and 92$^{\circ}$ (see inset of Figure \ref{st_obs}), respectively, relative to the Sun-Earth line. This positioning allowed the two spacecrafts to provide near-limb views of Earth-directed CMEs. On the other hand, the location of AR 11261 was close to the solar disk center (between 30$^{\circ}$to 35$^{\circ}$ west) relative to Earth, offering reliable magnetic field data for the active region from the Heliospheric Magnetic Imager or HMI instrument \citep{schou2012hmi}. This provides a unique opportunity to study the co-temporal evolution of reconnection flux, estimated from on-disk SDO data, and the speed of the associated eruption, starting from the initiation height at lower corona as observed from near-limb observations by STEREO.

\begin{figure*}[!t]
\centering
\includegraphics[width=0.8\linewidth]{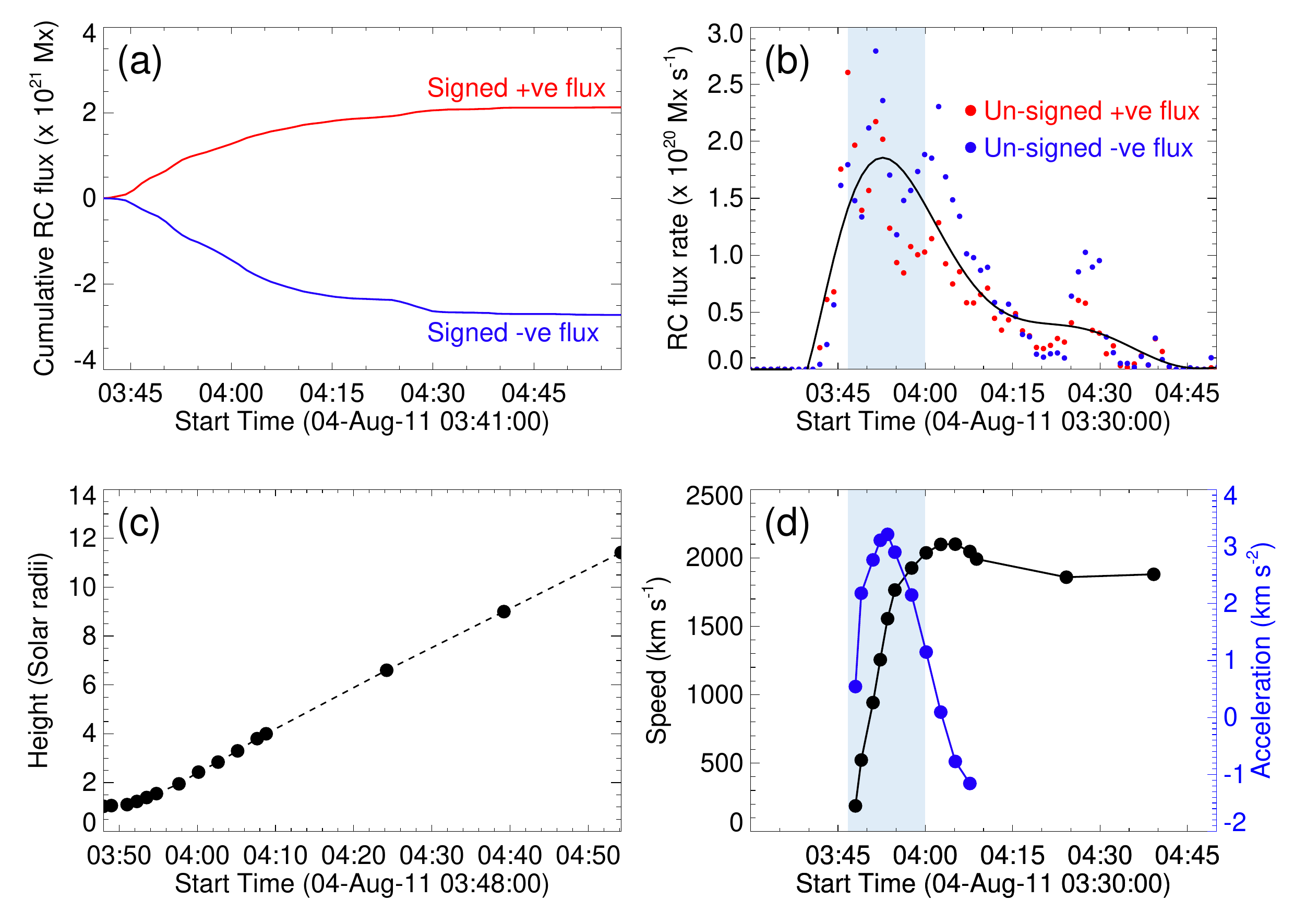}  
\caption{(a) Time profile of signed cumulative reconnection flux integrated over the positive (red) and negative (blue) magnetic polarities underlying the flare ribbons. (b) Time profile of unsigned instantaneous reconnection flux integrated over the positive (red solid circles) and negative (blue solid circles) magnetic polarities underlying the flare ribbons. The black solid line is the fitted profile over all the data points, including the red and blue solid circles. (c) Height time plot of the CME leading edge measured from STEREO EUVI, COR1 and COR2 for the same event as shown in Figure \ref{st_obs}. (d) Velocity (in black) and acceleration (in blue) profiles of the CME leading edge derived from Panel (c).}
\label{fig:combined_obs}
\end{figure*}

We analyze the kinematics of the eruption associated with the M-class flare on 2011 August 04, by tracking the CME leading-edge height as observed by STEREO-A. We use three instruments from the Sun Earth Connection Coronal and Heliospheric Investigation (SECCHI; \cite{howrard2008secchi}) onboard STEREO-A to track the early evolution of the Earth-directed CME eruption (see Figure \ref{st_obs}). The instruments include the Extreme Ultraviolet Imager EUVI ($< 1.7 R_\odot$), \cite{wuelser2004euvi} and the inner coronagraph COR1 ($1.3- 4 R_\odot$) and COR2  \citep{thompson2003cor1}.

We follow the method by \citep{Kazachenko2023Database}, to estimate the temporal profile of reconnecting flux. The observations of the flare ribbons are obtained from the Atmospheric Imaging Assembly (AIA; \cite{lemen2012aia}) onboard SDO in 1600 \AA\ channel. The associated full-disk magnetograms are obtained from the HMI instrument. During the recurrent eruptive M-class flare in AR 11261 on 2011 August 4, we identify the area swept out by the flare-ribbons observed in AIA 1600 \AA\ images as shown in Figure \ref{rec_mag}. The half of the unsigned magnetic flux underlying the area associated with the brightening observed in AIA 1600 \AA\ are recorded as the temporal profile of the re-connection flux. The kinematics of the CME eruption and the temporal variation in associated reconnection flux are shown in Figure \ref{fig:combined_obs}.

\subsection{Correlation between MFR acceleration and reconnection flux}

The panels (c) and (d) of Figure \ref{fig:fluxrate} show scatter plots of the MFR acceleration ($a_r$) versus the reconnection rate ($\phi_{RC}$) at different times during the evolution of the homologous synthetic eruptions. Before the eruption, the acceleration exhibits a 
monotonic relationship with the reconnection rate for the eruptive events. Notably, Pearson’s correlation coefficients before the eruption for the two events are 0.58 and 0.81, respectively, 
indicating a positive correlation between the reconnection rate and the acceleration. A similar qualitative relationship is also observed in the case of observational data (see panel (d) of Figure \ref{fig:combined_obs}) 
where the acceleration of the erupting filament and the associated reconnection flux simultaneously increase, as indicated by the cyan shaded region. The correlation coefficients for this observed event before and after the eruption are 0.90 and 0.98, respectively. Furthermore, note that, in the lower corona, where the influence of solar wind is negligible, our results show that the speed of the flux rope decreases following the eruption, 
heralding the end of the acceleration phase. The deceleration of the CME is mainly because of adiabatic expansion and the resulting cooling of the magnetic cloud in a relatively large domain that we have used. This is illustrated in panels (e) \& (f) of Figure \ref{fig:combined_sim} 
from our simulation model, which closely resembles panel (d) of Figure \ref{fig:combined_obs} based on our analysis of observational data. Similar findings have also been reported in \cite{Sarkar_2019}. \cite{Zhu2020} analyzed 60 CME-flare events and found correlation coefficients ranging from 0.68 to 0.99, indicating a strong correlation between acceleration and reconnection rate. Although our observational event is not included in their list, it also exhibits a similar correlation.
Due to observational constraints (limited time cadence of STEREO EUVI), it was challenging to gather more data points during the initial acceleration phase of the eruption. The numerical analysis from our simulation complements the above-mentioned observational results, allowing us to determine the temporal evolution of the reconnection rate and acceleration and their correlation.

\section{Summary and Discussion}

Our fully compressible MHD simulation using the Pencil Code presented here produces homologous CMEs due to the repeated formation and partial eruption of unstable flux ropes during the sustained emergence of highly twisted magnetic flux rope mimicking a newly formed coronal flux rope. Our results show that, with the continuous emergence of flux, a new current sheet (CS) forms above the same polarity inversion line (PIL) after the completion of a previous eruption, leading to a new eruption. This recurring formation and disruption of the coronal current sheet, driven by the ongoing emergence of flux, results in homologous eruptions.

Our simulation began with the quasi-static emergence of a twisted flux rope at the lower boundary into a pre-existing coronal arcade. We found that the flux rope initially settles into a quasi-static rise phase, with an underlying sigmoid-shaped current layer developing beneath the flux rope. This current layer is likely the site for the formation of thin current sheets and magnetic reconnection \citep{titov2007,aulanier2005,savcheva2012}. The reconnection in the current layer effectively adds twisted flux to the magnetic flux rope, allowing it to rise quasi-statically before undergoing a partial eruption. Subsequent flux emergence forms a new current channel, which exhibits similar behavior and erupts after the bodily emergence stops, or $v_0=0$.

We estimate by calculating the twist of the field line between the line tied ends that the first eruption remains stable to helical kink instability, while the second eruption is kink unstable. The subsequent CME, in our simulation, has a slightly lower speed and kinetic energy. Initially, the reconnection rate increases, followed by a gradual decrease, after each eruption. These trends are consistent with the observation of an eruptive event. Our simulation indicates that the reconnection process potentially influences the evolution of the magnetic flux rope. The reconnection rate shows a monotonic variation with the acceleration, especially before the eruption in our simulation.
The importance of the present work lies in performing an MHD simulation to produce homologous CMEs and thereafter finding the correlation between the reconnection rate and CME acceleration as a function of time between pre to post-eruptive stage and further validating this relation using an observed event. We choose a event that appears in the database of both STEREO (as limb event) and SDO (as on-disk event) spacecrafts so as to enable us to estimate both speeds and reconnection flux as accurately as possible using these instruments. Despite the fact that the driving of the MHD simulation was not tailored to match the observed flare and the associated CME, a similar pattern has been detected in both cases between CME acceleration and reconnection rate, (see panel (b) of Figure \ref{fig:combined_obs}), suggesting a consistent underlying process. 
The observed decrease in reconnection rate post-peak might be tied to the dynamic changes within the reconnection sheet during an eruption. As the reconnection sheet disintegrates, accurately measuring and tracking the magnetic field lines through the current sheet becomes challenging. This complexity impacts the calculated values of RC flux, reflecting the real-time physical changes occurring during an eruption. Moreover, these findings align with observational data, further validating the temporal relation between the reconnection rate and the onset of eruptions (panels b-d of Figure~\ref{fig:combined_obs}). 
This highlights the importance of considering reconnection rate during the rising phase of MFR evolution in studying the dynamics of solar eruptions and magnetic reconnection events. The correlation between RC flux rate and acceleration could therefore be a generic property of solar eruptions.

Although our numerical simulation model does not include the solar wind, the inclusion of optically thin radiative cooling and explicit coronal heating makes it more reasonable than previously published homologous solar eruption models \citep{chatterjee2013, bian2022}. We account for localized heating due to the formation and dissipation of current sheets in the corona, as well as the redistribution of heat through field-aligned conduction. The lower boundary of our simulation domain is positioned at the coronal base rather than the photosphere, and therefore, our model does not simulate prominence formation.

Although we employ an analytical configuration of the magnetic flux rope and include various physical processes in our model, this approach may not fully represent the realistic driving conditions of the solar eruptions. To address the complexity and evolution of the magnetic configuration in a real scenario, a detailed and accurate description of the evolving surface magnetic field is essential. Our future goal is to conduct detailed analysis of solar eruptions that are directly driven or constrained by photospheric magnetograms.

\section*{Acknowledgment}
The work was supported by Indo-US Science and Technology Forum (IUSSTF/JC-113/2019). SSM acknowledge NOVA HPC and VINO server of IIA used to perform the numerical simulation and analysis. P.C. acknowledges the DiRAC Data Intensive service (DIaL2) at the University of Leicester (project id: dp261), managed by the University of Leicester Research Computing Service and CSD3 service at the University of Cambridge on behalf of the STFC DiRAC HPC Facility (www.dirac.ac.uk). The DiRAC service was funded by BEIS, UKRI and STFC capital funding and STFC operations grants. DiRAC is part of the UKRI Digital Research Infrastructure. P.C. also acknowledges the allocation of computing
resources at the PDC Center for High Performance Computing at KTH in Stockholm, funded by the National Academic Infrastructure for Supercomputing in Sweden. RS acknowledges support from the Research Council of Finland Grant 350015. Additionally, authors thank the NASA SDO team  for providing valuable HMI and AIA data. SDO is a mission under NASA's Living with a Star program.  Open access is funded by Helsinki University Library.

\bibliographystyle{aasjournal} 
\bibliography{references}{} 

\end{document}